\documentclass[%
 reprint,
superscriptaddress,
 amsmath,amssymb,
 aps,
 prl,
]{revtex4-2}
\usepackage[T1]{fontenc}
\usepackage{amsfonts}
\usepackage{graphicx}
\usepackage{dcolumn}
\usepackage{bm}
\usepackage{braket}
\usepackage{booktabs}
\usepackage{xcolor}
\usepackage{siunitx,xfrac}
\usepackage[normalem]{ulem}  
\DeclareSIUnit{\sqrthz}{\sqrt{\hertz}}
\DeclareSIUnit{\dBm}{dBm}
\DeclareSIUnit{\dBc}{dBc}
\DeclareSIUnit{\dBi}{dBi}
\DeclareSIUnit{\Vcm}{(V/cm)}
\DeclareSIUnit{\cmV}{(cm/V)}
\DeclareSIUnit{\uVcm}{(\micro V/cm)}
\DeclareSIUnit{\micron}{\micro m}
\DeclareSIUnit{\uVm}{(\micro V/m)}
\DeclareSIUnit{\Vm}{(V/m)}
\DeclareSIUnit{\mVm}{(mV/m)}

\newcommand{\kb}[2]{|#1 \rangle\langle #2|}
\begin{document}

\preprint{APS/123-QED}

\title{Spin-Wave Quantum Computing with Atoms in a Single-Mode Cavity}

\author{Kevin C. Cox}
\email[Corresponding author: ]{kevin.c.cox29.civ@army.mil}
\affiliation{
 DEVCOM Army Research Laboratory, Adelphi, MD 20783 USA
}%

\author{Przemyslaw Bienias}
\affiliation{Joint Quantum Institute and Joint Center for Quantum Information and Computer Science, NIST/University of Maryland, College Park, Maryland 20742, USA}

\author{David H. Meyer}
\affiliation{
 DEVCOM Army Research Laboratory, Adelphi, MD 20783 USA
}%

\author{Paul D. Kunz}
\affiliation{
 DEVCOM Army Research Laboratory, Adelphi, MD 20783 USA
}%

\author{Donald P. Fahey}
\affiliation{
 DEVCOM Army Research Laboratory, Adelphi, MD 20783 USA
}%

\author{Alexey V. Gorshkov}
\affiliation{Joint Quantum Institute and Joint Center for Quantum Information and Computer Science, NIST/University of Maryland, College Park, Maryland 20742, USA}

\date{\today}

\begin{abstract}
We present a method for network-capable quantum computing that relies on holographic spin-wave excitations stored collectively in ensembles of qubits.  We construct an orthogonal basis of spin waves in a one-dimensional array and show that high-fidelity universal linear controllability can be achieved using only phase shifts, applied in both momentum and position space.  Neither single-site addressability nor high single-qubit cooperativity is required, and the spin waves can be read out with high efficiency into a single cavity mode for quantum computing and networking applications. 
\end{abstract}

\maketitle

Future quantum computers will likely be most useful when connected together into a quantum network, much like classical computers.  Quantum processors based on ions and superconducting qubits are advancing in their capability to perform high-fidelity operations at an intermediate scale \cite{pagano_quantum_2020, harrigan_quantum_2021}.  However, networking is still a serious challenge in these and other quantum processors, since qubits must be coupled to an optical communication channel with high efficiency and fast de-multiplexing.  Here, we discuss an alternative scheme, using quantum information stored as one-dimensional spin-wave holograms in an ensemble of qubits  \cite{wesenberg_quantum_2009,vasilyev_quantum_2010}.

Quantum memories based on collective excitations, often called spin waves, stored in ensembles of atoms have made significant progress \cite{choi_entanglement_2010,yang_multiplexed_2018, pu_experimental_2017, parniak_wavevector_2017}, and are an excellent candidate for quantum networking applications because of their potential for high capacity and strong coupling to a single optical mode.  However, spin-wave memories and processors have not yet demonstrated the capability for universal quantum computing, since the spin waves do not naturally interact strongly with one another.  In this work, we propose a method for universal spin-wave quantum computing using a high capacity spin-wave register inside of a single-mode optical cavity.  

We first describe the spin-wave computation basis in a one-dimensional lattice and the physical apparatus consisting of two atomic ensembles coupled to a single optical cavity mode.  Second, we describe the operations that enable universal quantum computing within the ensemble memory, specifically discussing the implementation of in-situ linear-optical quantum computing with spin waves.  We evaluate the fundamental performance limitations of the required operations and briefly discuss how they may be used to create high-speed entanglement generation in a quantum network.  This Letter is accompanied by a joint Article \cite{cox_linear_2021} describing how the general concept may be implemented, more specifically, in an experiment consisting of laser-cooled rubidium atoms coupled to an optical ring cavity.  The joint Article also presents a method to achieve universal quantum information processing using a single ensemble (instead of two, as discussed here), as well as a proposal for deterministic continuous-variable quantum computing using spin-wave cluster states.  We assert that our proposal may be realized using current experimental techniques, operating at performance levels already demonstrated in laboratory settings.

This Letter builds upon a significant body of recent work in many-body cavity quantum electrodynamics \cite{mcconnell_entanglement_2015, hacker_cavity_2017, dogra_formation_2018, bentsen_photon-mediated_2019, kroeze_sign-changing_2019, kesler_continuous_2020,clark_observation_2020, niedenzu_unraveling_2020} and memory experiments showing that quantum information can be stored as holographic spin waves in a large group of qubits \cite{wesenberg_quantum_2009, vasilyev_quantum_2010, cox_spin-wave_2019, yang_multiplexed_2018,reim_multimode_2008,  lan_multiplexed_2009}  that can be transferred to a readout bus for retrieval \cite{simon_interfacing_2007, parniak_wavevector_2017, ranjan_multimode_2020, heller_cold-atom_2020}.  Related work has theoretically investigated quantum information processing with holographic spin waves in superconducting circuits \cite{wesenberg_quantum_2009}.     Holographic quantum information is continuous, amenable to powerful and efficient quantum error correction \cite{gottesman_encoding_2001}, and can have collectively-enhanced qubit-light coupling \cite{cox_spin-wave_2019}.  However, no proposals have presented a method for efficient, universal quantum computation, because collective atomic ensembles have negligible atom-atom interactions, which are typically used for gate operations.   Phase modulation of spin waves has been used to observe interference between two holographic profiles \cite{parniak_quantum_2019}, but spatial phase shifts alone are not sufficient to create arbitrary linear unitary operations, a condition we refer to as linear controllability, that is required for universal quantum processing \cite{reck_experimental_1994}.  
By combining spatial phase gradients with a collective cavity dressing operation, implementing phase shifts in momentum space, we establish a method for full linear controllability, and therefore universal linear optical quantum computing.  In Ref.~\cite{cox_linear_2021}, we extend this proposal to a deterministic continuous-variable quantum computing paradigm that also relies on linear controllability.  

\begin{figure}[t]
\centering
\includegraphics[width=\columnwidth]{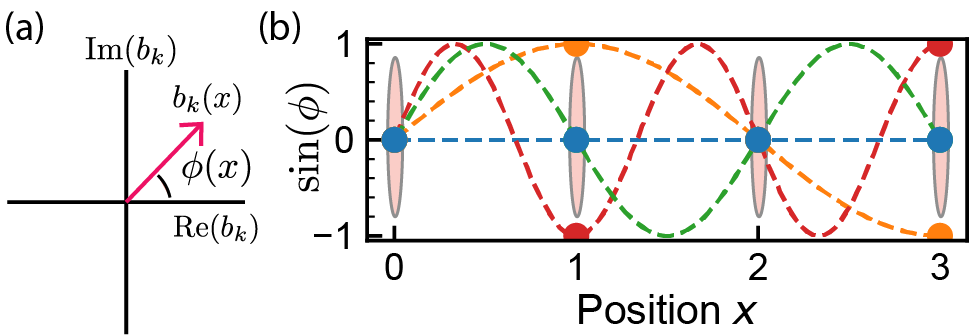}
\caption{Momentum basis. a) Spin-wave eignenstates are characterized by an excitation amplitude $b_k(x)$ with spatially-dependent phase $\phi(x)$,  $b_k(x) = \frac{1}{\sqrt{M}} e^{i\phi(x)} = \bra{ggg...} \hat{a}_x \hat{b}_k^\dagger \ket{ggg...}$.   b) Momentum eigenstates with $M=4$ sites, labeled by momentum number $k$ ranging from $k=0$ to $k=3$ (blue, orange, green, and red, respectively).  Dashed lines are guides for the eye, with physical phases represented by solid points.}
\label{fig:fig1}
\end{figure}

\begin{figure}[t]
\centering
\includegraphics[width=\columnwidth]{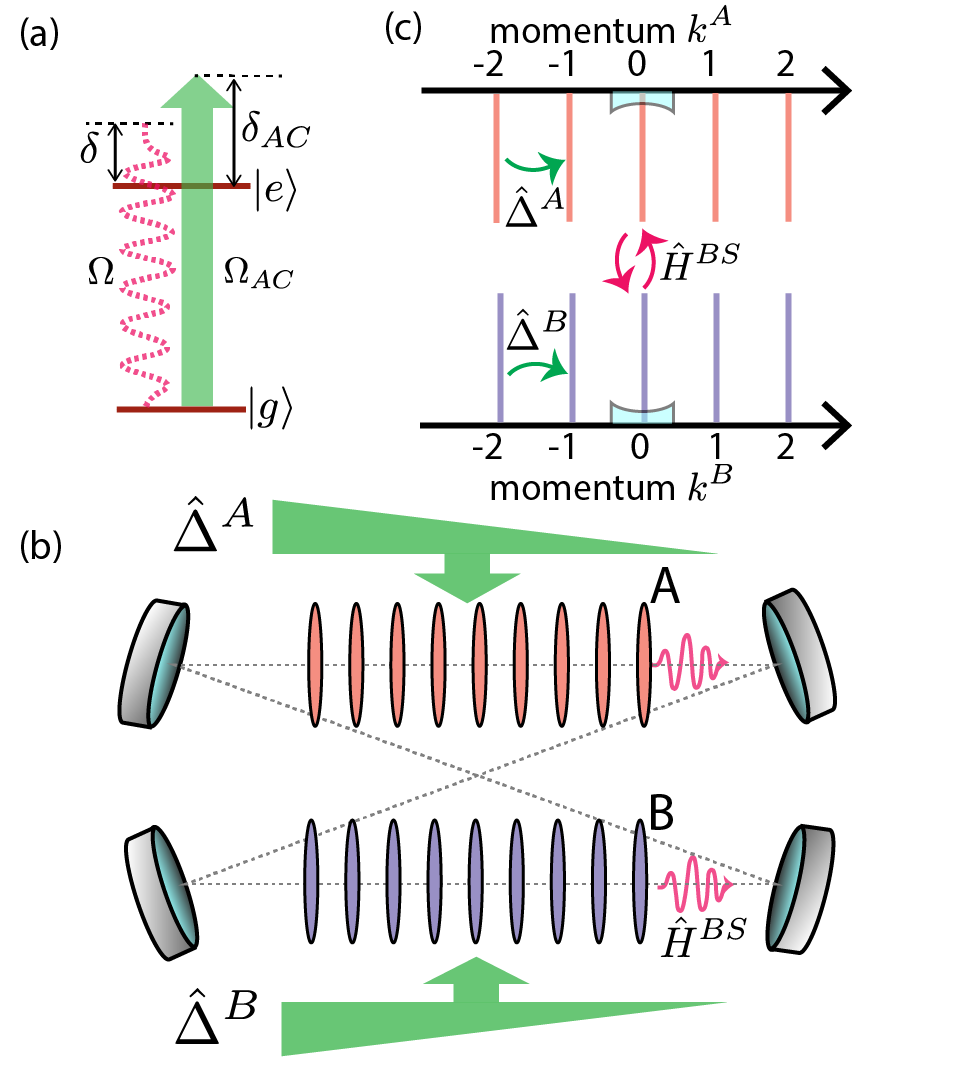}
\caption{ (a) Level diagram. The atomic transition is tuned off resonance from an optical cavity mode (pink, collective vacuum Rabi frequency $\Omega$, detuning $\delta$) and optical potential (green, Rabi frequency $\Omega_{AC}$, detuning $\delta_{AC}$).  (b) Physical system and two-ensemble approach.  Two ensembles of atoms interact with a single cavity mode but with independent cavity couplings and optical gradients $\hat{\Delta}$. (c) Connectivity diagram.  The momentum shifts $\hat{\Delta}$ and cavity beamsplitter $\hat{H}_{BS}$ allow full connectivity for spin waves.  $k$ is defined modulo $M$ so that $k=-1$ is equivalent to $k = M-1$. }
\label{fig:fig2}
\end{figure}

We consider two ensembles, each consisting of $N$ atoms in a one-dimensional array of $M$ sub-ensembles with $n = N/M$ atoms per site, shown in Figs.~\ref{fig:fig1} and \ref{fig:fig2}.  We label experimental parameters and operators with a superscript $A$ or $B$ when referring to one specific ensemble.  The atoms interact with a single optical mode, taken here to be defined by an optical ring cavity \footnote{A travelling wave, or ring, cavity is required to distinguish spin-wave momenta corresponding to clockwise and counter-clockwise cavity emission}.  The atom-light coupling is described by the Jaynes-Cummings coupling parameter $g$ for a single atom.  Optical excitations in an ensemble are collective in nature, being stored in an equal (up to phases) superposition of all $N$ atoms.   The lowering operator that describes a collective excitation at array site $x$ is defined as,
\begin{equation}
    \hat{a}_x = \frac{1}{\sqrt{n}}\sum_{l=0}^{n-1} \kb{g_l}{e_l},
\end{equation}
where we sum over all atoms $l$ at site $x$ and $\ket{g_l}$ and $\ket{e_l}$ are the ground and excited states of the two-level atoms.  

In this work, we are interested in the corresponding momentum states, where excitations are stored in an equal superposition of all the lattice sites, with spatially dependent phase.  The orthogonal set of momentum operators are defined over the full set of $M$ array sites:
\begin{equation}
    \hat{b}_k = \frac{1}{\sqrt{M}} \sum_{x=0}^{M-1} e^{i 2 \pi x k / M} \hat{a}_x.
\end{equation}
The integer index $0 \leq k \leq M-1$ delineates the orthogonal basis of momentum operators.  The zero-momentum excitations, created by the $\hat{b}^\dagger_0$ operator acting on the ground state $\ket{ggg...}$, are simply an equal superposition of all atoms, with common phase [see Fig.~\ref{fig:fig1}(b)].   Working with (including initializing and reading out) excitations in the $\hat{b}_k$ modes is experimentally advantageous because it relaxes the requirement for single-site addressability while retaining the full $M$-mode Hilbert space. We will show that any mode $\hat{b}_k$ can be efficiently read out into the cavity mode by using a momentum shift operation.

If the number of excitations at every site is small compared to $n$, $\hat{b}_k$ is equivalent to a canonical photon lowering operator in mode $k$ by the Holstein-Primakoff approximation \cite{holstein_field_1940}. In this work, we assume this approximation to be valid.  Spin-wave computing is also likely to be possible outside of the linear regime, but this is beyond the scope of this initial work.  As such, the spin-wave operators $\hat{b}_k$ define a set of $M$ independent bosonic modes, analogous to $M$ optical channels.  Unlike an optical system, the spin-wave excitations are stationary, stored as patterns in the large ensemble of atoms.  We show that linear optical quantum computing may be performed in this $M$-mode system by describing how to perform arbitrary linear unitary operations.

First, we describe the atom-cavity system and physical operations.  A level diagram is shown in Fig. \ref{fig:fig2}(a).  A single-mode optical cavity (pink) is detuned by $\delta$ from the atomic transition.  We use an off-resonant cavity interaction to apply phase shifts in momentum space, to the $k=0$ spin wave, and a free-space potential gradient (green) to apply phase shifts in position space.  Here we consider free-space phase shifts applied with an off-resonant optical potential, but a magnetic field or another potential could also be used.  

To easily achieve linear  controllability, we propose a two-ensemble implementation, shown in Fig.~\ref{fig:fig2}(b).  Two $M$-site ensembles of atoms, labeled A and B, couple to a single running-wave cavity mode.  Each ensemble has cavity coupling that may be turned on and off independently \cite{cox_linear_2021}.  Also, potential gradients (green arrows) may be applied to the two ensembles separately.  Importantly, the two-ensemble approach is not a strict requirement for spin-wave computing.  An alternative experimental realization using one ensemble is derived in Ref.~\cite{cox_linear_2021}.  However, the two-ensemble approach lends itself to simple operations and experimental implementation, as discussed below.  

The optical gradients (green in Fig.~\ref{fig:fig2}) create a standard AC-Stark shift potential with spatially-dependent Rabi frequency $\Omega_{AC}(x)$, described (up to an additive constant) by the Hamiltonian 
\begin{equation}
    \hat{H}_\Delta =\sum_{x=0}^{M-1}-\hbar \frac{\Omega_{AC}^2(x)}{2 \delta_{AC}} \hat{a}^\dagger_x  \hat{a}_x.
\end{equation}
If $\Omega_{AC}^2(x) = \alpha x$ is linear, it imposes a phase gradient (momentum shift) to all spin waves stored in either ensemble A or B. For application time $\tau =(4 \pi |\delta_{AC}|)/(\alpha M)$,  the resulting unitary is (assuming negative detuning $\delta_{AC} = -|\delta_{AC}|$) 
\begin{equation}
    \hat{\Delta} = \sum_{x=0}^{M-1} \left[\sum_{l=0}^{n-1}\left(\ket{g_{l,x}}\bra{g_{l,x}}+ e^{2 \pi i x/M} \ket{e_{l,x}}  \bra{e_{l,x}}\right)\right],
\end{equation}
up to a global phase.  $\hat{\Delta}$ translates all spin-wave excitations by one momentum unit: $\hat \Delta \hat b_k^\dagger = \hat b_{k+1}^\dagger \hat \Delta$, where $k+1$ is evaluated modulo $M$.  Graphically, the effect of this operation is shown in the connectivity diagram in Fig.~\ref{fig:fig2}(c).  By applying $\hat{\Delta}^A$ and $\hat{\Delta}^B$, an excitation in the A and B ensembles may be translated relative to one another in momentum space.  Shifting a mode $\hat{b}_k$ by the maximal amount $k \rightarrow k + M/2$, can be achieved in constant time (independent of $M$), since the application time $\tau$ scales inversely to $M$.  Any two spin waves may be simultaneously aligned with zero momentum, $k^{A}=0$ or $k^{B}=0$, that collectively interacts with the cavity mode bus.

The spin-waves aligned with $k=0$ interact with the cavity with strength characterized by the collective vacuum Rabi splitting $\Omega = 2 g \sqrt{N}$, where $g$ is the Jaynes-Cummings atom-cavity coupling parameter.  We assume that $g^{A}$ and $g^{B}$ may be switched on and off independently \cite{cox_linear_2021}.  This allows the application of phase shifts to $k^A = 0$ and $k^B=0$ spin waves and also implementation of a beamsplitter.

If a cavity coupling is turned on, for either ensemble A or B, in the limit of large detuning $|\delta| \gg \Omega$, a  cavity-dressing Hamiltonian results \cite{noauthor_suplementary_nodate}:
\begin{equation}
    \hat{H}_0^k = -\hbar \frac{\Omega^2}{4\delta} \hat{b}^\dagger_0 \hat{b}_0.
\end{equation}
This cavity coupling, in which all atoms couple symmetrically, interacts only with the $k=0$ spin wave.

If both cavity couplings $A$ and $B$ are simultaneously turned on, a beamsplitter Hamiltonian results instead:
\begin{equation}
    \hat{H}^{BS} = -\hbar \frac{\Omega^2}{4\delta} (\hat{b}^{\dagger A}_0 + \hat{b}^{\dagger B}_0)(\hat{b}^A_0 + \hat{b}^B_0). 
\end{equation}
By applying $\hat{H}^{BS}$ for various times, we implement an arbitrary 2-mode beamsplitter.  As shown in the connectivity diagram of Fig.~\ref{fig:fig2}(c), we can use $H^{BS}$ and $\hat{\Delta}$ to achieve the desired splitting ratio between any mode  in ensemble $A$ with any mode in ensemble $B$ by phase shifting any pair of spin waves into the zero-momentum mode that interacts with the cavity.  The beamsplitter operation $H^{BS}$ and the spin-wave phase shift operation $\hat{H}_0^k$ are primary elements of this approach that extend beyond previous spin-wave processing procedures.

Most importantly, the two-mode beamsplitters $H^{BS}$, in combination with the momentum displacement operators $\hat{\Delta}$ and spin-wave phase shifts $\hat{H}_0^k$, create a fully connected graph of spin-wave modes [Fig.~\ref{fig:fig2}(c)], sufficient for linear controllability in the register of $2M$ modes \cite{reck_experimental_1994}. This construction of efficient linear controllability is the major result of this work, since linear controllability and memory readout are precisely the two requirements for universal linear optical quantum computing \cite{knill_scheme_2001}.  This approach  has a further advantage experimentally in that any of the $M^2$ possible beamsplitters may be accomplished in constant (independent of $M$) time, although there is a disadvantage that multiple beamsplitters cannot be run in parallel. 



Initialization of a large number of spin-wave quanta is another important task for a spin-wave processor, required for linear optical quantum computing.  Many approaches may be taken.  One may simply use an appropriate single-photon source \cite{ripka_room-temperature_2018,ornelas-huerta_-demand_2020} to seed excitations into the system.  Many quantum memory systems using alkali atoms instead create single excitations using a heralded process \cite{simon_interfacing_2007}.  In this heralded process, discussed in more detail in Ref.~\cite{cox_linear_2021}, the rate is limited by the linewidth of the D2 transition ($\Gamma \sim 2 \pi \times 10~$MHz) and the efficiency is limited to a small number $p$, so that the two-photon excitation rate $\propto p^2$ is minimized.  Nonetheless, one may expect to create hundreds or thousands of quanta at milli-second timescales \cite{cox_linear_2021}.
\begin{figure}[t]
\centering
\includegraphics[width=\columnwidth]{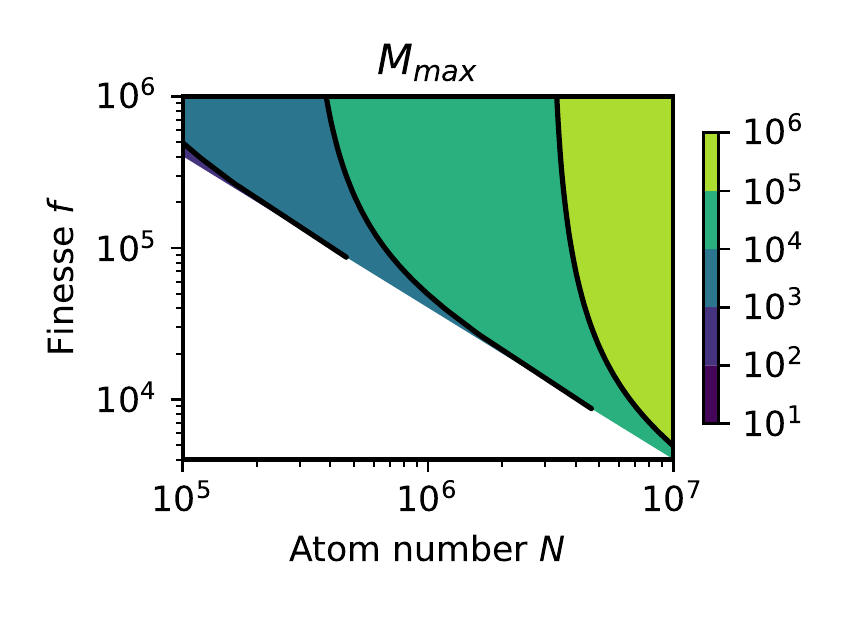}
\caption{Maximum processor capacity $M_{max}$ versus finesse $f$ and total atom number $N$.  $M_{max}$ is defined as the maximal capacity $M$ that leads to a beamsplitter error of less than $10^{-3}$ for a single-atom cooperativity of $C = 10^{-4}f$. 
}\label{fig:fig3}
\end{figure}

We briefly consider the leading fundamental sources of error in the spin-wave processor.  In particular, we consider three primary sources of error that limit the fidelity of the beamsplitter operation $\hat{H}^{BS}$:  free-space emission, cavity emission, and atomic saturation.  First we estimate the error $E_M$ in a beamsplitter due to atomic saturation.  Assuming that order $M$ excitations are spread approximately equally within the modes, the leading error can be calculated using the Holstein-Primakoff relations \cite{holstein_field_1940} and scales as $E_M \sim 1/n^2 = M^2/N^2$ \cite{noauthor_suplementary_nodate}.  Next, consider the error $E_{FS}$ due to free-space atomic emission.  Since the beamsplitter takes time $T \sim \delta/\Omega^2$ and the emission probability is $\Gamma$ times $T$, this error is of order
\begin{equation}
    E_{FS} \sim \frac{\Gamma \delta}{\Omega^2} = \frac{\delta}{\kappa N C}
\end{equation}
for a cavity with FWHM linewidth $\kappa$ and collective cooperativity $NC = \Omega^2/(\kappa \Gamma)$.  $E_{FS}$ is minimized at smaller detunings $\delta$.  The error due to cavity emission $E_c$, on the other hand is
\begin{equation}
    E_c \sim \frac{\kappa}{\delta}.
\end{equation}
$E_c$ improves with larger detuning.  These errors are discussed in more detail in the Supplemental Material \cite{noauthor_suplementary_nodate}.  The minimal total error $E = E_M + E_{FS}+E_{c}$ is optimized at a detuning $\delta_{opt} = \kappa \sqrt{NC}$, where $E_{FS} \sim E_{c} \sim 1/\sqrt{NC}$.  In Fig.~\ref{fig:fig3}, we plot the maximum capacity $M_{max}$ at which the total error $E$ is less the $10^{-3}$ for a cavity with single-atom cooperativity $C = 10^{-4}f$ and finesse $f$.  In the white region of the plot, the error $E$ is above $10^{-3}$ even for $M=1$.  This region corresponds to $2/\sqrt{NC} > 10^{-3}$.  Since the maximum possible detuning is given by the free spectral range (FSR) $\delta \sim $ FSR  $=\kappa f$, large finesse $f > 10^{3}$ is also required to achieve $E < 10^{-3}$.  Importantly, for less-than state-of-the-art experimental values of $N \sim 5\times 10^5$ and $f \sim 5\times 10^5$, the error is $E<10^{-3}$ even with a capacity of over $M = 10^3$.  For larger atom numbers, $M_{max}>10^4$ is possible.  Additional technical sources of error are discussed in Ref. \cite{cox_linear_2021}.

Overall, the spin-wave processor is well-suited to perform linear-optical computing for several reasons. First, each spin wave may be read out into one mode using the cavity bus \cite{simon_single-photon_2007}.  Recent atom-cavity quantum memory experiments have demonstrated efficiencies of over 80\% for this collection process, and read-out efficiencies approaching unity are feasible \cite{simon_interfacing_2007,yang_efficient_2016}.  Second, unlike photonic platforms, the atomic qubits are stationary and can have a coherence time up to one million times longer than the length of an operation \cite{zhao_long-lived_2009, yang_efficient_2016}, allowing more sophisticated computing than has yet been demonstrated using photons.  Third, and most important, the atomic system may integrate universal processing into a high-capacity quantum memory, capable of forming the building block of a high-speed quantum network.  Our approach to universal quantum information processing appears particularly well-suited to the task of efficient long-distance entanglement distribution, i.e., a quantum repeater. Not only does the ensemble-cavity coupling provide an efficient light-matter interface and multiplexing capacity, but we have now shown that it has the ability to internally perform beamsplitter operations. By performing these operations on the spin waves, we avoid the losses associated with transmitting photons through discrete optical elements. 

In the joint Article \cite{cox_linear_2021}, we discuss a specific implementation of this proposal for linear controllability using a laser-cooled and trapped alkali atoms in a Raman configuration.  We further extend upon this work by proposing a continuous variable scheme that relies on spin-wave cluster states, that may be useful for achieving fault tolerance with favorable scaling \cite{menicucci_universal_2006,gu_quantum_2009,bourassa_blueprint_2021}.

There is still a long research path to arrive at quantum information processors with wide ranging practical utility beyond the scientific research community.  Development of new paradigms and platforms for quantum information processing is a key area that will require significant future work.  Holographic spin-wave quantum information processing using collectively-enhanced ensembles can be a key technique in the future.  The spin-wave technique combines universal quantum computing capabilities, potentially with high fidelity, into compact and networkable quantum memories that can lead to increased-performance quantum repeaters and networks.

\begin{acknowledgments}
The authors would like to thank Fredrik Fatemi for useful discussions and guidance. P.B.~and A.V.G.~acknowledge funding by ARO MURI, DARPA SAVaNT ADVENT, AFOSR MURI, AFOSR, NSF PFCQC program, DoE ASCR Accelerated Research in Quantum Computing program (award No.~DE-SC0020312), the DoE ASCR Quantum Testbed Pathfinder program (award No.~DE-SC0019040), and U.S.~Department of Energy Award No.~DE-SC0019449.  
\end{acknowledgments}

\label{app:ExpDetails}

\bibliography{bib1}

\end{document}